\documentclass[useAMS,usenatbib]{mn2e}
\usepackage{graphicx,times}
\usepackage{bm}
\newcommand{\EQ}{\begin{equation}}
\newcommand{\EN}{\end{equation}}
\newcommand{\EQA}{\begin{eqnarray}}
\newcommand{\ENA}{\end{eqnarray}}
\newcommand{\mean}[1]{\overline{ #1}}

\newcommand{\meanB}{\overline{B}}

\newcommand{\meanU}{\overline{U}}

\newcommand{\meanEMF}{\bmath{\cal E}}
\newcommand{\meanBB}{\bmath{\overline{{B}}}}
\newcommand{\meanJJ}{\bmath{\overline{{J}}}}
\newcommand{\meanUU}{\bmath{\overline{{U}}}}
\newcommand{\ko}{k_0}

\newcommand{\uu}{\bmath{u}}
\newcommand{\BB}{\bmath{B}}
\newcommand{\JJ}{\bmath{J}}
\newcommand{\jj}{\bmath{j}}
\newcommand{\AAA}{\bmath{A}}
\newcommand{\aaaa}{\bmath{a}}
\newcommand{\bb}{\bmath{b}}

\newcommand{\FF}{\bmath{F}}

\def\Rm{R_\mathrm{m}}

\def\half{{\textstyle{1\over2}}}

\def\onethird{{\textstyle{1\over3}}}

\newcommand\B{B}

\newcommand{\crit}{_\mathrm{c}}

\newcommand{\M}{_\mathrm{m}}
\newcommand{\etat}{\eta_\mathrm{t}}
\newcommand\deriv[2]{\displaystyle\frac{\partial #1}{\partial #2} }
\newcommand\sfrac[2]{{\textstyle{\frac{#1}{#2}}}}

\newcommand{\Veff}{U}
\newcommand{\Beq}{\B_\mathrm{eq}}

\newcommand{\gam}{\gamma}

\newcommand{\w}{\widehat{W}}
\newcommand{\vp}{\widehat{V}}
\newcommand{\e}{\epsilon}
\newcommand{\Co}{C_0}
\newcommand{\Cop}{C^{'}_{0}}
\newcommand{\Cl}{C_1}
\newcommand{\Clp}{C^{'}_{1}}

\newcommand{\s}{\,{\rm s}}

\newcommand{\cm}{\,{\rm cm}}

\newcommand{\kms}{\,{\rm km\s^{-1}}}

\newcommand{\mkG}{\,\mu{\rm G}}

\newcommand{\p}{\,{\rm pc}}

\newcommand{\yr}{\,{\rm yr}}

\newcommand{\RU}{R_U}
\newcommand{\Ral}{R_\alpha}
\newcommand{\Rw}{R_\omega}

\title{Galactic dynamos supported by magnetic helicity fluxes}
\author[S.~Sur, A.~Shukurov and K.~Subramanian]%
{Sharanya Sur,$^{1}$\thanks{E-mail: sur@iucaa.ernet.in (SS);
anvar.shukurov@ncl.ac.uk (AS); kandu@iucaa.ernet.in (KS)}
Anvar Shukurov$^{1,2}$ and Kandaswamy Subramanian$^{1}$\\
$^{1}$Inter-University Centre for Astronomy and
        Astrophysics,  Post Bag 4, Ganeshkhind, Pune 411 007, India\\
$^{2}$School of Mathematics and Statistics, Newcastle University, Newcastle
        upon Tyne, NE1 7RU, U.K.}

\date{}
\begin{document}

\pagerange{\pageref{firstpage}--\pageref{lastpage}} \pubyear{2006}

\maketitle

\begin{abstract}
We present a simple semi-analytical model of nonlinear, mean-field galactic
dynamos and use it to study the effects of various magnetic helicity fluxes.
The dynamo equations are reduced using the `no-$z$' approximation to a
nonlinear system of ordinary differential equations in time; we
demonstrate that the model reproduces accurately earlier results, including
those where nonlinear behaviour is driven by a magnetic helicity flux. We
discuss the implications and interplay of two types of magnetic helicity flux,
one produced by advection (e.g., due to the galactic fountain or wind) and the
other, arising from anisotropy of turbulence as suggested by \citet{VC01}. We
argue that the latter is significant if the galactic differential rotation is
strong enough: in our model, for $\Rw\la-10$  in terms of the corresponding
turbulent magnetic Reynolds number. We confirm that the intensity of gas
outflow from the galactic disc optimal for the dynamo action is close to that
expected for normal spiral galaxies. The steady-state strength of the
large-scale magnetic field supported by the helicity advection is still weaker
than that corresponding to equipartition with the turbulent energy. However,
the Vishniac-Cho helicity flux can boost magnetic field further to achieve
energy equipartition with turbulence. For stronger outflows that may occur in
starburst galaxies, the Vishniac-Cho flux can be essential for the dynamo
action. However, this mechanism requires a large-scale magnetic field of at
least $\simeq1\mkG$ to be launched, so that it has to be preceded by a
conventional dynamo assisted by the advection of magnetic helicity by the
fountain or wind.
\end{abstract}
\label{firstpage}

\begin{keywords}
magnetic fields -- turbulence -- ISM: magnetic fields -- galaxies: magnetic
fields -- galaxies: ISM
\end{keywords}

\section{Introduction}
Conservation of magnetic helicity has recently been recognized as a key
constraint on the evolution of cosmic magnetic fields, especially those
produced by the large-scale dynamo action. Any large-scale magnetic field (if
not supported by external currents) must have both poloidal and toroidal parts
in order to mitigate diffusive losses; therefore, any self-sustained
large-scale magnetic field must have non-zero magnetic helicity. One of
consequences of the helicity conservation is the suppression of the
$\alpha$-effect in a medium of high electrical conductivity. Since the
large-scale magnetic field has a non-zero helicity in each hemisphere through
the mutual linkage of poloidal and toroidal fields, the dynamo must be
producing small-scale helical magnetic fields with the opposite sign of the
magnetic helicity. In several closure models \citep*{PFL76,KR82,GD94,BF02},
this leads
to a suppression of the dynamo action due to the current helicity of the
growing small-scale magnetic field. The suppression can be catastrophic in the
sense that the steady-state energy density of the mean magnetic field is as
small as $\Rm^{-1}$ times the kinetic energy density, where $\Rm\gg1$ is the
magnetic Reynolds number. Thus, efficient action of the mean-field dynamo
requires that the magnetic helicity due to the small-scale magnetic field is
removed from the dynamo active region
\citep{BF00,BF01,KMRS00}.

Several mechanisms have been suggested to produce the required helicity flux,
including the anisotropy of the turbulence produced by differential rotation
\citep{VC01,SB04}, and the non-uniformity of the $\alpha$-effect
\citep{KMRS00,KMRS02}. More recently, \citet{SSSB06} suggested a simpler
mechanism involving advection of small-scale magnetic fields (together with
their helicity) by an outflow  from the dynamo-active region, e.g., the
galactic fountain or wind in the case of spiral galaxies. We
explore these effects using a simple model where dynamo equations are reduced
to a system of ordinary differential equations, which captures the salient
features of the nonlinear dynamo action and helicity evolution. The simplicity
of the model allows us to gain deeper insight into the role and interaction of
various mechanisms, and to more extensively explore the parameter space. We
consider both an advective helicity flux and that arising from the combined
action of velocity shear due to differential rotation and anisotropy of the
turbulence produced by it. We demonstrate the efficiency of the galactic
dynamo action provided it is facilitated by the helicity fluxes.

\section{Helicity balance and mean-field dynamos}\label{HBMFD}
The generation of a large-scale magnetic field by small-scale random motions
is described by the mean-field electrodynamics \citep{KR80}, where relevant
physical quantities are split into  mean and fluctuating parts, e.g.,
$\BB=\meanBB+\bb$ for magnetic field and $\AAA=\mean{\AAA}+\aaaa$ for the
vector potential, where overbar denotes relevant averaging and $\mean{\bb}=0$,
$\mean{\aaaa}=0$. This results in the mean-field dynamo equation
\EQ
\frac{\partial \meanBB}{\partial t}=
\nabla\times(\meanUU\times\meanBB+\meanEMF-\eta\meanJJ)\,,
\label{fullset1flux}
\EN
where $\meanUU$ is the mean velocity field, $\eta$ is the Ohmic magnetic
diffusivity, and $\JJ=\nabla\times\BB/\mu_0$ (for the vacuum magnetic
permeability we assume $\mu_0=1$ hereafter). Furthermore,
\[
\meanEMF \equiv \mean{\uu \times \bb} = \alpha\meanBB -\etat\meanJJ
\]
is the mean turbulent electromotive force (emf) (assuming isotropic
turbulence), with $\alpha$ and $\etat$ the turbulent transport coefficients
responsible for the $\alpha$-effect and turbulent magnetic diffusion,
respectively, and $\uu$ is the small-scale velocity field. The dependence of
$\alpha$ on $\meanBB$ and $\bb$ is a subject of ongoing research, and our
knowledge of this is based on limited closure models and numerical
simulations. In some closure models, such as EDQNM \citep*{PFL76}
and the $\tau$-approximation \citep*{BF02,RKR03,BS05b}, the effect
of the small-scale magnetic field on the $\alpha$-effect is described by
\citep{BS05a}
\begin{equation}\label{aquench}
\alpha = \alpha_\mathrm{K}+\alpha\M\,,
\end{equation}
where $\alpha_\mathrm{K}$ represents the kinetic $\alpha$-effect,
$\alpha_\mathrm{K}=-\sfrac13\mean{\tau\uu\cdot\nabla\times\uu}$, and $\alpha\M
= \sfrac13\rho^{-1}\mean{\tau\jj\cdot\bb}$ is the magnetic contribution to the
$\alpha$-effect, with $\rho$ the fluid density and $\tau$ the correlation time
of the turbulent velocity field $\uu$ (assumed to be short). Further,
$\eta_{\rm t} =\sfrac13 \mean{\tau \uu^2}$.  It is not clear how general is
the form (\ref{aquench}). Using a low Reynolds number approximation, where
unambiguous analysis is feasible, \citet*{SSB07} argue that
Eq.~(\ref{aquench}) is acceptable, with $\alpha_\mathrm{K}$ modestly affected
by magnetic field.

To constrain $\alpha\M$, we use the helicity conservation equation written in
terms of the helicity density $\chi$ of the small-scale magnetic field,
defined in a gauge-invariant form in terms of the number density of the links
of $\bb$. Since the small-scale magnetic field has finite correlation length,
one can meaningfully introduce its {\em helicity density\/} and then derive
its transport equation in the form \citep{SB06}
\EQ
\frac{\partial \chi}{\partial t} + \nabla\cdot\FF
= -2\meanEMF\cdot\meanBB-2\eta\overline{\jj\cdot\bb}\,,
\label{finhel}
\EN
where $\FF$ is the helicity flux density (specified below) and
$\jj=\nabla\times\bb$. For practical purposes, $\chi$ is approximately equal
to $\mean{\aaaa\cdot\bb}$ given that $\aaaa$ is defined in the Coulomb gauge.

In the steady state, Eq.~(\ref{finhel}) yields $\meanEMF\cdot\meanBB =
-\half\nabla\cdot\FF -\eta\mean{\jj\cdot\bb}$. For $\FF=0$, it follows that
$\meanEMF\cdot\meanBB\to0$ as $\eta\to0$ under reasonable assumptions about
the spectrum of current helicity $\overline{\jj\cdot\bb}$ (specifically that
it peaks at a scale independent of $\eta$ or at least at such a scale that
$\eta\mean{\jj\cdot\bb}\to0$ as $\eta\to0$). Hence the component of the emf
parallel to the mean magnetic field vanishes, and so the dynamo becomes
inefficient. This catastrophic quenching of the $\alpha$-effect in highly
conducting medium is, however, avoided if $\FF\neq0$. The simplest
contribution to the flux,
\begin{equation}\label{FcU}
\FF = \chi\meanUU\,,
\end{equation}
arises from the net effect of advection of magnetic fields by a velocity field
$\meanUU$ directed away from the dynamo active region, as suggested by
\citet{SSSB06}. We also include another helicity flux proposed by
\citet{VC01}.

In order to relate $\alpha\M$ to $\chi$, we argue that the main contribution
to $\alpha\M$ comes from the integral scale of the turbulence $l_0=2\pi/\ko$
\citep{SSSB06}, so that $\mean{\jj\cdot\bb}\simeq
l_0^{-2}\mean{\aaaa\cdot\bb}$ and then
\begin{equation}\label{amchi}
\alpha\M \simeq\onethird\tau \,\frac{1}{l_0^2} \,\frac{\chi}{\rho}\,.
\end{equation}
To justify this relation, we note that the spectral density of
$\alpha\M\propto\mean{\tau\jj\cdot\bb}$ differs from that of
$\chi\simeq\mean{\aaaa\cdot\bb}$ by a factor $\tau_k k^2$. When the spectral
density of $\chi$ decreases with $k$ fast enough, the spectrum of $\alpha\M$
decreases with $k$ as well, and then both $\chi$ and $\alpha\M$ are dominated
by the integral scale, $2\pi/k_0$, which is the meaning of Eq.~(\ref{amchi}).
For example, for the Kolmogorov spectrum $M_k=k^{-1}b_k^2\propto k^{-5/3}$, we
have $b_k\propto k^{-1/3}$ and $\tau_k\propto k^{-2/3}$, so that the spectrum
of $\chi$ is $a_k b_k\propto k^{-5/3}$, that of the current helicity is
$j_k b_k\propto k^{1/3}$, and $\tau_k j_k b_k\propto k^{-1/3}$ for $\alpha\M$.
Then both $\alpha\M$ and $\chi$ are dominated by the integral scale. These
arguments are similar to those presented in Sect.~10.IV of \citet*{ZRS83}.
Analytical \citep{B03} and numerical results \citep{BS05b} suggest that the spectrum of the
current helicity in fact decreases as $k^{-2/3}$; this lends further support
to Eq.~(\ref{amchi}).

Introducing a reference magnetic field strength $\Beq^2\equiv\rho \mean{u^2}$
and defining the magnetic Reynolds number as $\Rm=\etat/\eta$, we rewrite
Eq.~(\ref{finhel}), using Eqs~(\ref{FcU}) and (\ref{amchi}), as
\EQ
{\partial\alpha\M\over\partial t}=-\frac{2\etat}{l_0^2}
\left({{\meanEMF\cdot\meanBB}\over \Beq^2}+{\alpha_{\rm m}\over \Rm}\right)
-\nabla\cdot\left(\alpha_{\rm m}\meanUU\right).
\label{fullset2flux}
\EN
This equation should supplement the mean-field dynamo equations to provide our
model of nonlinear dynamo action. Since galactic discs are thin, it suffices
to consider a one-dimensional model, retaining only the $z$-derivatives of the
variables \citep*{RSS88}. In terms of cylindrical coordinates $(r,\phi,z)$, we
consider a mean flow consisting of differential rotation and vertical
advection, with ${\meanUU} = (0, \meanU_\phi,\meanU_z)$.

Altogether, the system of nonlinear mean-field dynamo equations has the
dimensionless form
\begin{equation}
\label{mfBrnoz1}
{\partial\meanB_r\over\partial t}=-{\partial\over\partial
z}\left(\RU\meanU_z\meanB_r+\Ral\alpha\meanB_\phi\right)
+{\partial^2\meanB_r\over\partial z^2}\,,
\label{mfBphinoz1}
\end{equation}
\begin{equation}\label{Bphiao}
{\partial\meanB_\phi\over\partial t}=
{\Rw\meanB_r}-\RU{\partial\over\partial z}\left(\meanU_z\meanB_\phi\right)
+{\partial^2\meanB_\phi\over\partial z^2}\,,
\end{equation}
\begin{equation}
{\partial\alpha_{\rm m}\over\partial t}=-C\left(\alpha\meanB^2
        -\Ral^{-1}\mean{\JJ}\cdot\mean{\BB}+{\alpha_{\rm m}\over \Rm}\right)
-\RU{\partial\over\partial z}\left(\alpha\M\meanU_z\right),
\label{fullset2fluxnoz1}
\end{equation}
where, retaining only the $z$-derivatives,
\begin{equation}\label{JB}
\mean{\JJ}\cdot\mean{\BB}=
                \meanB_\phi{\partial\meanB_r\over\partial z}
                -\meanB_r{\partial\meanB_\phi\over\partial z}\,,
\end{equation}
and we have neglected the term proportional to $\Ral$ in Eq.~(\ref{Bphiao})
thus adopting the $\alpha\omega$-dynamo approximation for convenience; this
term can easily be restored. Here the time and length units are $h^2/\etat$
and $h$, respectively, with $h$ the semi-thickness of the disc; the unit of
$\alpha$ is denoted $\alpha_0$, and that of $\mean{U}_z$ is $U_0$; magnetic
field is measured in the units of $B_\mathrm{eq}$. We have introduced
dimensionless parameters defined as
\EQ
\RU={{\Veff_0}h \over\eta_{\rm t}}\,,\hfil
\Rw={Gh^2\over\etat}\,,\hfil
\Ral={{\alpha_0}h \over\eta_{\rm t}}\,,\hfil
C=2\left(\frac{h}{l_0}\right)^2,
\EN
where $G=r\,d\Omega/dr$ and we have neglected $\mean{B}_z$ in comparison with
the other two components of the mean magnetic field, which is appropriate in a
thin disc \citep{RSS88}. We shall also use the dynamo number defined as
$D=\Ral\Rw$.

\section{Reduced dynamo equations} \label{noz}
We reduce the dynamo equations to a dynamical system, where all the variables
are functions of time alone, by using the `no-$z$' approximation
\citep{SM93,M95,P01}. Specifically, the $z$-derivatives of all the variables
are replaced by appropriate division by the disc semi-thickness,
$\partial/\partial z \to 1/h$ and $\partial^2/\partial z^2\to-1/h^2$. The
field components ${\meanB}_r, {\meanB}_\phi$ appearing in the resulting
ordinary differential equations can be thought of as representing either the
corresponding mid-plane values or  vertical averages. By its nature, the
no-$z$ approximation captures the quadrupole modes of the dynamo which are
dominant in a thin disc.

The current helicity density of the large-scale field
$\mean{\JJ}\cdot\mean{\BB}$ vanishes in the no-$z$ approximation -- see
Eq.~(\ref{JB}). Therefore, we calculated it using a  perturbation solution of
one-dimensional, kinematic mean-field dynamo equations as described in
Appendix~\ref{appen}:
\begin{equation}\label{JdB}
\mean{\JJ}\cdot\mean{\BB} \approx
                        -\sfrac{3}{8}(-\pi D)^{1/2}\meanB_r\meanB_\phi\,.
\end{equation}
When applying this estimate to study nonlinear solutions, we recall that
$D\propto\alpha=\alpha_\mathrm{K}+\alpha\M$. Furthermore, we follow
\citet{P01} who suggested that the accuracy of the no-$z$ approximation can be
improved by using
\EQ \label{nozd}
\alpha \to {2 \over \pi}\alpha, \quad
{\partial^2 \over \partial z^2} \to -{\pi^2 \over 4}\,\frac{1}{h^2}\;.
\EN
These rescalings are consistent with the $z$-dependence of $\mean{\BB}$
obtained in Appendix~\ref{appen}. Derivatives similar to
$\partial(\alpha\meanB_\phi)/\partial z$ have to be treated cautiously in the
no-$z$ approximation since this term is positive near $z=0$ and negative near
$z=1$ (note that  $\alpha=0$ at $z=0$ and $\meanB_\phi=0$ at $z=1$). The main
contribution to the dynamo action comes from the vicinity of $z=0$ where
$|\meanB_\phi|$ is maximum, so that the part of the above term important for
the dynamo action is $\meanB_\phi\partial\alpha/\partial z$ \citep{S96}, where
$\partial\alpha/\partial z>0$ near $z=0$ in galactic discs. This suggests that
the no-$z$ approximation of this term is $\alpha\meanB_\phi/h$ rather than
$-\alpha\meanB_\phi/h$.

The resulting reduced Eqs~(\ref{mfBrnoz1})--(\ref{fullset2fluxnoz1}) are given
by
\EQA
\label{mfBrnoz2}
{d\meanB_r\over dt}&=&
-{2\over\pi}\Ral\left(1+{\alpha_{\rm m}}\right)\meanB_\phi
-\left(\RU+{\pi^2\over4}\right)\meanB_r\,, \\
\label{mfBphinoz2}
\frac{d\meanB_\phi}{dt}
                &=& \Rw\meanB_r-\left(\RU+{\pi^2\over4}\right)\meanB_\phi\,,\\
\nonumber
\frac{d\alpha\M}{dt}
        &=&-\RU\alpha\M-C\left[(1+\alpha\M)(\meanB_r^2+\meanB_\phi^2\right.)\\
        &&\mbox{}+(1+\alpha\M)^{1/2}\left.
                \frac{3(-\pi D)^{1/2}}{8\Ral}\meanB_r\meanB_\phi
        +\frac{\alpha\M}{\Rm}\right]\;,
\label{fullset2fluxnoz2}
\ENA
where $\alpha_\mathrm{K}=\alpha_0$ is absorbed into $\Ral$, so that
$\alpha=1+\alpha\M$ in terms of dimensionless variables, and $\mean{U}_z$ can
now be put equal to unity. Note that $D$ must be replaced by $D(1+\alpha\M)$
when Eq.~(\ref{JdB}) is used in the helicity evolution equation because $D$ in
that equation has the meaning of the instantaneous dynamo number.

Parameter values typical of the Solar neighbourhood of the Milky Way are as
follows. The scale height of the ionized gas layer is $h\simeq500\p$, the
integral scale of interstellar turbulence is $l_0\simeq100\p$, which yields
$C=2(h/l_0)^2\simeq50$. With the turbulent magnetic diffusivity
$\etat\simeq10^{26}\cm^2\s^{-1}$ and $U_0=1\kms$ \citep{SSSB06}, we obtain
$\RU\approx1.5$; we consider a range $\RU=0\mbox{--}2$. Furthermore, we use
standard values of the dynamo parameters, $\Ral=1$ and $\Rw=-15$. These
estimates are often adopted as typical of spiral galaxies in general. However,
many galaxies rotate faster than the Milky Way, and even within the Milky Way
the angular velocity of rotation rapidly increases towards the centre.
Likewise, the other parameters vary broadly between galaxies and across a
given galaxy. Hence, any identification of the above parameter ranges as
generally applicable to spiral galaxies should be treated with caution.

To illustrate the accuracy of the no-$z$ approximation, we calculate the
critical dynamo number (corresponding to $\partial\meanBB/\partial t=0$) from
Eqs~(\ref{mfBrnoz2}) and (\ref{mfBphinoz2}) with $\alpha\M=0$ and $\RU=0$ as
$D\crit\approx-(\pi/2)^5\approx-9.6$, as compared to the more accurate values
(obtained from a numerical solution of the boundary value problem)
$D\crit\approx-8$ and $-11$ for $\alpha_\mathrm{K}=\sin\pi z$ and $z$,
respectively \citep{RSS88}. In the next section we demonstrate that relevant
features of the numerical solutions of the nonlinear equations
(\ref{mfBrnoz1})--(\ref{fullset2fluxnoz1}), obtained by \citet{SSSB06}, are
accurately reproduced as well.

\section{The effects of the advective flux}\label{ssb1}
We solved numerically Eqs~(\ref{mfBrnoz2})--(\ref{fullset2fluxnoz2}) firstly
in order to verify the accuracy of the no-$z$ approximation, and then to study
the nonlinear evolution of the dynamo. The initial conditions used are
\EQ \label{bcondsnoz}
\meanB_r=0\,, \quad
\meanB_\phi=10^{-3}\,, \quad
\alpha\M=-10^{-3}
\quad\mbox{ at $t=0$}.
\EN
The choice $\alpha\M|_{t=0}=0$ does not affect the results.

\begin{figure} \begin{center}
\includegraphics[width=1.0\columnwidth,angle=0,bbllx=18bp,
bblly=264bp,bburx=591bp,bbury=708bp,clip=true]{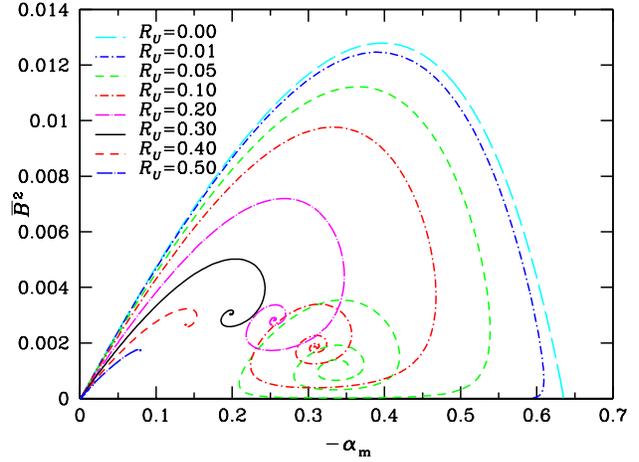}
\end{center}\caption[]{
The phase portrait of the dynamical system (\ref{mfBrnoz2})--(\ref{bcondsnoz})
in the $(\meanB^2,-\alpha\M)$-plane with and without vertical advection,
$0\leq\RU\leq0.5$. Parameter values used here are $\Ral=1$, $\Rw=-15$, $C=50$,
$\Rm=10^5$, the values of $\RU$ are shown in the legend, and the time span is
$t\leq 100 h^2/\etat$.
} \label{phplcomb}
\end{figure}

In Fig.~\ref{phplcomb} we show the phase portrait of the system in the
$(\meanB^2,-\alpha\M)$-plane, obtained for various values of $\RU$. For weak
advection ($\RU\la0.01$), $-\alpha\M$ monotonically increases with time to
reduce the net $\alpha$-effect, so that the dynamo eventually becomes
subcritical and magnetic energy reduces to negligible values. The dynamo
action can resume again as soon as $\alpha\M$ will have decayed together with
the large-scale magnetic field and $\alpha=\alpha_\mathrm{K}+\alpha\M$ becomes
supercritical again. Without the advective flux of helicity, the time scale
for such a recovery of the dynamo would be of the order of the Ohmic diffusion
time, as controlled by the term $\alpha\M/\Rm$ on the right-hand side of
Eq.~(\ref{fullset2fluxnoz2}). However, if $\RU$ is of order unity, the
relaxation time of the magnetic helicity becomes significantly shorter because
of the advection represented by the first term on the right-hand side of
Eq.~(\ref{fullset2fluxnoz2}), and magnetic field grows again to exhibit
nonlinear oscillations before it approaches a steady state. The time scale of
the oscillations is long at about $2\times10^{10}\yr$ for $\RU=0.05$. The
magnitude of $\meanB^2$ in the steady state first grows with $\RU$, but then
reaches a maximum at $\RU\approx0.3$. Stronger advection affects the dynamo
adversely by removing $\meanBB$ too fast.

\begin{figure} \begin{center}
\includegraphics[width=1.0\columnwidth,angle=0,
bbllx=18bp,bblly=260bp,bburx=586bp,bbury=710bp]{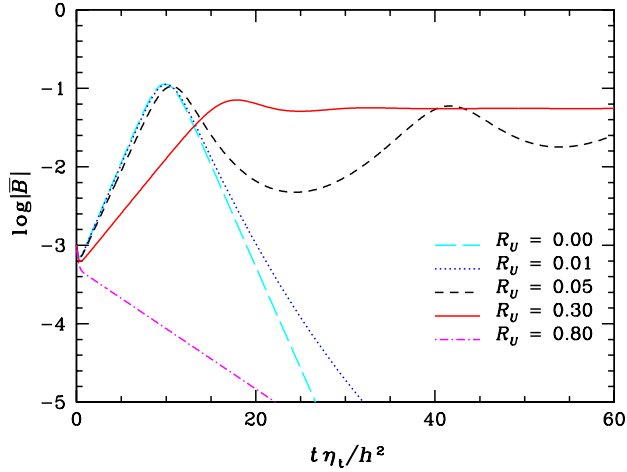}
\end{center}\caption[]{
The evolution of the magnetic field strength for various magnitudes of the
vertical advective flux shows that magnetic field eventually decays for both
small and large values of $\RU$. Parameter values are the same as in
Fig.~\ref{phplcomb}. Time is given in the units of the magnetic diffusion time
across the gas layer, about $7.5\times10^8\yr$.
} \label{Bevolve}
\end{figure}

The variation of magnetic field strength with time is also shown in
Fig.~\ref{Bevolve}. The initial exponential growth of the magnetic field is
catastrophically quenched in the absence of the advective flux, and the field
decays at about the same rate as it grew (long-dashed). However, even a
moderate advective flux ($\RU=0.3$, solid) compensates the catastrophic
quenching allowing the magnetic field to reach a steady-state value of about
$0.1\Beq$. For stronger advection ($\RU\geq0.8$, see the dash-doted curve),
the dynamo action is suppressed again since the mean-field is removed too
rapidly from the dynamo-active region.

The above results agree quite accurately with those obtained by \citet{SSSB06}
who solved the differential equations
(\ref{mfBrnoz1})--(\ref{fullset2fluxnoz1}). This confirms the applicability of
the no-$z$ approximation to nonlinear dynamo equations. In addition, this
indicates that our estimate of $\meanJJ\cdot\meanBB$ in Eq.~(\ref{JdB}) is
appropriate.

To clarify further the effect of the advection on the dynamo, we consider the
steady-state solution, $d/dt=0$ in
Eqs~(\ref{mfBrnoz2})--(\ref{fullset2fluxnoz2}), where we introduce the
critical dynamo number $D\crit$ such that, in the steady state,
\begin{equation}\label{amcr}
\left[D(1+\alpha\M)\right]_{\mbox{steady state}}=D\crit\,.
\end{equation}
Then Eqs~(\ref{mfBrnoz2}) and (\ref{mfBphinoz2}) yield
\begin{equation}\label{DcrBr}
D\crit=-\frac{\pi}{2}\left(\RU+\frac{\pi^2}{4}\right)^2\;,
\quad
\meanB_r=-\left(\frac{2D\crit}{\pi}\right)^{1/2}\,\frac{\meanB_\phi}{\Rw}\;,
\end{equation}
the first of which shows that the critical dynamo number is affected by the
advection; in this sense, the outflow hinders the dynamo action, as might be
expected.The second of this relations shows that $|\meanB_r|\ll|\meanB_\phi|$
if $|\Rw|\gg1$. Then Eq.~(\ref{fullset2fluxnoz2}) yields, neglecting
$\meanB_r^2$ in comparison with $\meanB_\phi^2$,
\begin{equation} \label{Bnozss}
\meanB^2\approx \frac{\xi}{C}
\left({D \over D\crit}-1\right)
\left(\RU+{C\over\Rm}\right),
\end{equation}
where
\begin{equation}\label{xiii}
\xi=\frac{1}{1-\sfrac38\sqrt2}\approx2\,.
\end{equation}
It is clear that $\meanB\propto\Rm^{-1/2}$ for $\RU=0$, which is the case of
catastrophic $\alpha$-quenching \citep[with $\meanB$ perhaps exhibiting
long-term variations around this level --][]{BS05c}. However, the magnetic
field strength is approximately proportional to $\RU^{1/2}$ for $\Rm\gg1$ and
small $\RU$. On the other hand, $|D\crit|$ increases with $\RU$, so that there
 is an optimal value of $\RU$ providing maximum field strength. For $D=-15$,
the steady state is nontrivial ($\meanB^2>0$) for $\RU\la0.6$, with $\meanB^2$
being maximum for $\RU\approx0.3$. In the optimal case, $\RU=0.3$, we obtain
$\meanB\approx0.1$, where we recall that the unit magnetic field corresponds
to equipartition between magnetic and turbulent energy densities. We note that
Eq.~(\ref{Bnozss}) is similar to Eq.~(10) of \citet{SSSB06}, but here we use
an arguably better estimate of $\meanJJ\cdot\meanBB$.

The steady-state magnetic field remains of order $C^{-1/2}\Beq\simeq0.1\Beq$,
with $C=2(h/l_0)^2\simeq50$, even in the presence of the advective flux.
Therefore, we consider additional helicity fluxes to examine if they can lead
to stronger magnetic fields such that $\mean{B}\simeq\Beq$.

\section{The Vishniac-Cho flux} \label{vcflux}
A flux of magnetic helicity discovered by \citet{VC01} relies on the
anisotropy of turbulence which is naturally produced by velocity shear due to
differential rotation \citep{SB04,BS05c}. Indeed, the regular velocity shear
$\meanU_y(x)$ produces an additional $y$-component of the turbulent velocity
via $\partial u_y/\partial t \simeq (\uu\cdot\nabla)\meanU_y$, so that the
resulting anisotropy, produced during one correlation time $\tau$, is
$u_y/u_x\simeq1+\tau\partial\meanU_y/\partial x$. The $x$ and $y$ directions
can be identified with the radial and azimuthal ones in the galactic disc, so
that $\partial\meanU_y/\partial x$ is replaced by $G=r\,d\Omega/dr$. The
anisotropic part of the turbulent velocity correlation tensor can be estimated
as $\mean{u_x u_y}\simeq\tau G \mean{u^2}$, where $\uu$ is the background
isotropic turbulent velocity field. A convenient expression for the flux of
Vishniac \& Cho has been obtained by \citet{BS05c} and \citet{SB06}. Using
Eq.~(12) of \citet{BS05c}, with their $C_\mathrm{VC}$ calculated using the
results of \citet{SB06}, we can represent the vertical flux of the magnetic
helicity of the small-scale magnetic field in the following dimensional form:
\begin{equation}\label{VCF}
F_z\simeq\sfrac12 (u\tau)^2 G(\meanB_r^2-\meanB_\phi^2)\;.
\end{equation}
This flux will add the following term to the right-hand side of
Eq.~(\ref{fullset2fluxnoz2}) written in the dimensionless form (and using the
no-$z$ approximation):
\EQ
{\partial\alpha_{\rm m}\over\partial t}=
\cdots-\deriv{F_z}{z}\,,
\qquad
\deriv{F_z}{z}\approx
-{\Rw\over\Ral}\left(\meanB_r^2-\meanB_\phi^2\right),
\label{vcflux_eq}
\EN
where dots denote the terms already included in Eq.~(\ref{fullset2fluxnoz2}).
Concerning the steady state, Eqs.~(\ref{amcr}) and (\ref{DcrBr}) still apply
and the steady-state strength of the mean field is given by
Eq.~(\ref{Bnozss}), but now with
\begin{equation}\label{BVC}
\xi=\frac{1}{1-\frac38\sqrt2+\Rw^2/(CD\crit)}
\end{equation}
instead of (\ref{xiii}). Since $D\crit<0$, the additional flux results in a
stronger steady-state magnetic field than that with advective flux alone --
cf.\ Eq.~(\ref{xiii}).

If $|\Rw|$ is large enough, the additional term can formally lead to a
singularity in $\meanB^2$ where the denominator in Eq.~(\ref{BVC}) vanishes
(and $\meanB^2<0$ for larger values of $|\Rw|$). The reason is that the
Vishniac-Cho flux produces $\alpha\M>0$ enhancing the hydrodynamic
$\alpha$-effect. If $\Rw^2$ is large enough in comparison with $C|D\crit|$, the
flux of Vishniac-Cho can lead to a dynamo action of its own, independently of
the kinetic $\alpha$-effect \citep{VC01}. As shown in Sect.~\ref{VCdynamo},
equations governing this regime are in fact linear in $\mean{\BB}$, hence
$\meanB^2$ grows exponentially as $t\to\infty$, which corresponds to the
singularity in Eqs~(\ref{Bnozss}) and (\ref{BVC}). Despite the formally linear
governing equations, this type of dynamo is essentially nonlinear and relies
on either a strong seed magnetic field or an earlier conventional dynamo
action to produce a strong magnetic field required to build up $\alpha\M$ at a
sufficiently short time scale (Sect.~\ref{VCdynamo}).

For the parameter values typical of the Solar neighbourhood, $C=50$ and
$D\crit=-10$, the effect of the additional helicity flux becomes significant
as $\Rw$ approaches about $-15$. For $|\Rw|\la15$, this effect does not lead
to independent dynamo action (see Sect.~\ref{VCdynamo}) but rather increases
the steady-state strength of the large-scale magnetic field to values close to
$\Beq$.

\begin{figure}
\begin{center}
\includegraphics[width=1.0\columnwidth,angle=0,
bbllx=34bp,bblly=180bp,bburx=563bp,bbury=718bp]{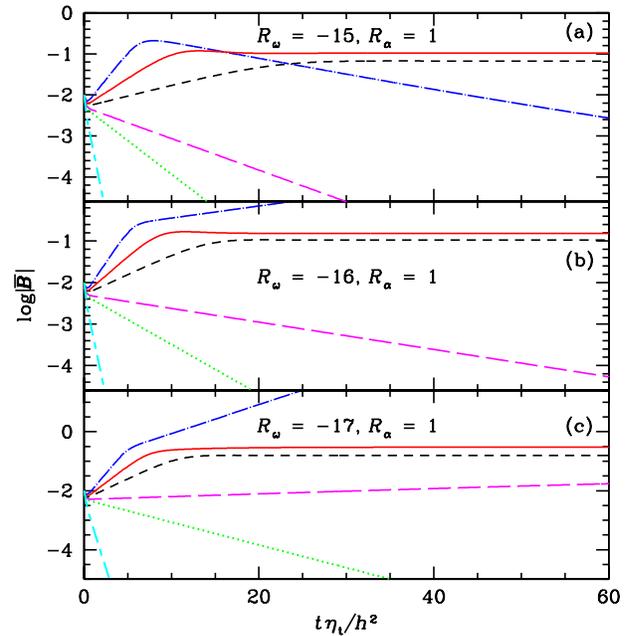}
\end{center}\caption[]{
Evolution of the magnetic field strength with both the advective helicity flux
and that of Vishniac \& Cho, flux for various magnitudes of the velocity
shear: (a)$\Rw=-15$, (b) $\Rw=-16$ and (c) $\Rw=-17$. The values of $\RU$
presented are: 0 (dash-dotted), 0.3 (solid), 0.5 (dashed), 0.8 (long-dashed),
1.0 (dotted) and 3.0 (short-long-dashed). The Vishniac-Cho helicity flux
supports the dynamo action for larger values of $\RU$: note decay at
$\RU=0.8$ in (a) but not in (c). The dynamo action is similarly facilitated
for small $\RU$, where solution with $\RU=0$ decays in (a) but not in (b)
and (c). The parameters used here are $C=50$, $\Rm=10^5$,
$\meanB_\phi|_{t=0}=10^{-2}$ and ${\meanB_r}|_{t=0}=0$, with the values of
$\Ral$ and $\Rw$ indicated in the legends.
}\label{VCB}
\end{figure}

With the addition of the Vishniac-Cho flux, we plot in Fig.~\ref{VCB} the
variation of the magnetic field strength driven by both helicity fluxes for a
particular value of $\Ral$ and varying $\Rw$, as obtained from a numerical
solution of Eqs~(\ref{mfBrnoz2})--(\ref{fullset2fluxnoz2}) with the
modification (\ref{vcflux_eq}). The useful contribution of the Vishniac-Cho
flux at larger values of $\RU$ is now evident. With the advective helicity
flux alone, magnetic field decays if $\RU$ is large enough; this is true with
the Vishniac-Cho flux added, but now for larger values of $\RU$. For example,
$\meanB$ decays for $\RU>0.8$ if $\Rw=-15$ (either with the flux of Vishniac
\& Cho -- Fig.~\ref{VCB}a, or without it -- Fig.~\ref{Bevolve}), but only for
$\RU>1$ if $\Rw=-17$ (Fig.~\ref{VCB}c). This is a result of the joint action
of the two mechanisms, the $\alpha\omega$-dynamo and that driven by the
Vishniac-Cho flux: both become stronger as $|\Rw|$ increases. For
$\Rw\leq-15.5$, the solutions for $\RU=0$ do not decay because the
Vishniac-Cho flux drives a dynamo of its own (the dash-dotted curves in
Fig.~\ref{VCB}b,c).

Thus, the advective helicity flux facilitates the dynamo action by alleviating
the magnetic helicity conservation constraint. On the other hand, it ensures
that the dynamo action is eventually saturated. On the contrary, the helicity
flux of Vishniac \& Cho, also mitigating the helicity constraint, can lead to
an unbounded growth of magnetic field and must be quenched by further physical
processes. The advective flux can ensure such a quenching for moderate values
of $|\Rw|$ (e.g., for $\RU=0.3$--0.8 in Fig.~\ref{VCB}).

\subsection{Dynamo action due to the Vishniac-Cho flux}\label{VCdynamo}
In this section we present an illustrative model of dynamo action driven by
the helicity flux of \citet{VC01} alone. For this purpose, we use
Eq.~(\ref{fullset2fluxnoz1}) with additional term (\ref{vcflux_eq}) but now we
do not employ the no-$z$ approximation, take  $\RU=0$ and neglect any
hydrodynamic $\alpha$-effect, $\alpha_\mathrm{K}=0$. To make the system easily
tractable, we make two key assumptions which can be justified a posteriori
using the solution obtained: we neglect $\meanB_r^2$ in comparison with
$\meanB_\phi^2$ in Eq.~(\ref{fullset2fluxnoz1}) and assume that
$\mean{\JJ}\cdot\mean{\BB}$ is negligible. Neglecting also $\alpha\M/\Rm$,
this yields
\[
\deriv{\alpha\M}{t}=-\alpha\M C \meanB_\phi^2-\deriv{F_z}{z}\,,
\]
where the dimensional form of $F_z$ is given in Eq.~(\ref{VCF}), and we use
dimensionless variables as defined in Sect.~\ref{HBMFD}.
Then Eqs~(\ref{mfBrnoz1})--(\ref{fullset2fluxnoz1}) reduce to
\begin{eqnarray}
\deriv{\meanB_r}{t}
        &=&-\deriv{}{z}(\alpha\M\meanB_\phi)+\deriv{^2\meanB_r}{z^2}\;,
        \label{VCdr}\\
\deriv{\meanB_\phi}{t}
        &=&\Rw\meanB_r+\deriv{^2\meanB_\phi}{z^2}\;,\label{VCdp}\\
\deriv{\alpha\M}{t}
        &=&-\alpha\M C \meanB_\phi^2+\Rw\deriv{\meanB_\phi^2}{z}
        \label{VCda}\;,
\end{eqnarray}
where we have put $\Ral=1$, which is equivalent to choosing the unit of
$\alpha\M$ to be $\etat/h$. The characteristic time of the variation of
$\alpha\M$, of the order of the turbulent time scale, is much shorter than
that of the magnetic field (see Sect.~\ref{disc}). Therefore, we can take
$\partial\alpha\M/\partial t=0$, which yields
\[
\alpha\M\simeq 2\frac{\Rw}{C}\deriv{\ln\meanB_\phi}{z}\,.
\]
Perhaps unexpectedly, Eqs.~(\ref{VCdr}) and (\ref{VCdp}) now become linear,
\begin{eqnarray}
\deriv{\meanB_r}{t}
        &=&-2\frac{\Rw}{C}\deriv{^2\meanB_\phi}{z^2}+\deriv{^2\meanB_r}{z^2}\;,
        \label{VCdr1}\\
\deriv{\meanB_\phi}{t}
        &=&\Rw\meanB_r+\deriv{^2\meanB_\phi}{z^2}\;,\label{VCdp1}
\end{eqnarray}
and have the following exact quadrupolar solution satisfying the vacuum boundary
conditions (\ref{BCq})
\[
\left(
\begin{array}{c}
\meanB_r\\
\meanB_\phi
\end{array}
\right)
=K
\left(
\begin{array}{c}
-\pi \kappa\sqrt{2/C}\\
1
\end{array}
\right)
\cos(\pi\kappa z)\exp{(\gamma t)}\,,
\]
where growing magnetic fields have
\[
\gamma=\pi\kappa\left(|\Rw|\sqrt{{2}/{C}}-\pi\kappa\right)\,,
\quad
\kappa=\sfrac12+n\,,
\quad
n=0,1,\ldots\,,
\]
and $K$ is an arbitrary constant. Given that $\Rw<0$ in galactic discs, the
dynamo produces a growing magnetic field if $\Rw<{\Rw}\crit$ with
\begin{equation}\label{Rwcrit}
{\Rw}\crit=-\pi\sqrt{C/2}\approx -8\,.
\end{equation}
As we argue below, this is a necessary but not sufficient condition for the
dynamo action of this type.

The assumptions made to derive this solution can now be verified:
$|\meanB_r/\meanB_\phi|=\pi/\sqrt{2C}\ll1$ for the lowest mode, and
$\mean{\JJ}\cdot\mean{\BB}\equiv0$  since $\meanB_r/\meanB_\phi$ is independent
of $z$ in this approximation.

An essential feature of this mechanism is that it cannot be launched unless
the large-scale magnetic field is strong enough: for $\meanB^2\ll1$, we have
$|\alpha\M|\ll1$ and the field must decay. For $\alpha\M$ to grow, its
variation rate $C\meanB^2$ should be larger than the magnetic diffusion rate
across the disc, $\pi^2/4$ of Eq.~(\ref{nozd}). This yields $\meanB\ga\pi/(2\sqrt{C})\approx0.2$,
where we recall that magnetic field is measured in the units of $\Beq$. For
galactic parameters, the minimum magnetic field is of order $1\mkG$. The
initial magnetic field needs to be even stronger if we take into account that
it first decays until $\alpha\M$ has grown enough.
These arguments equally apply to the helicity flux discussed by
\citet{KMRS00,KMRS02}.

Therefore, conditions for the dynamo action driven by the Vishniac-Cho alone are, firstly,
the inequality (\ref{Rwcrit}) and, secondly, the initial large-scale magnetic
field must be strong enough, say $\meanB^2|_{t=0}>B_0^2$, where
$B_0$ strongly depends on $\Rw$. As a result, the dynamo threshold ${\Rw}\crit$
is a function of the initial magnetic field; in this sense, Eq.~(\ref{Rwcrit})
is just a necessary condition. Numerical solution of the dynamo equations in the
no-$z$ approximation with $\alpha_\mathrm{K}=0$ shows that magnetic field decays
for $\Rw>-22$ for any initial magnetic field and grows for $\Rw<-23$
if $B_0>0.4$ and for $\Rw<-40$ if $B_0>0.1$.
This dynamo mechanism seems to be suitable for an additional
amplification of the large-scale magnetic field produced by the conventional
mean-field dynamo (assisted by the advective helicity flux as described in
Sect.~\ref{ssb1}) closer to equipartition with turbulent energy.

The Vishniac-Cho dynamo can be saturated by quenching the corresponding
helicity flux, with $\partial F_z/\partial z$ in Eq.~(\ref{vcflux_eq})
multiplied by $1/(1+\meanB^2)$. We have confirmed, using the no-$z$
approximation, that this indeed produces a steady state with $\meanB=O(1)$.
Curiously, the `standard' $\alpha$-quenching, with
$\alpha\M\to\alpha\M/(1+\meanB^2)$ in both Eqs~(\ref{VCdr}) and (\ref{VCda}),
cannot lead to a saturated state because of the trivial cancellation of the
quenching factor.

\section{Conclusions and discussion} \label{disc}
The simple model suggested above reproduces a feature of the mean-field dynamo
which has become well known: if the helicity of the small-scale magnetic field
cannot escape from the dynamo active region, the mean magnetic field
eventually decays to negligible levels being constrained by the conservation
of magnetic helicity. Correspondingly, Eq.~(\ref{Bnozss}) yields
$\meanB\simeq\Rm^{-1/2}\ll1$ for $\RU=0$. An outflow from the dynamo region
can prevent this catastrophic suppression of the dynamo as it carries away
small-scale magnetic fields together with their contribution to the total
magnetic helicity, giving breathing space to the large-scale magnetic field
\citep{SSSB06}. Thus, $\meanB\propto \RU^{1/2}(l_0/h)\Beq$ in Eq.~(\ref{Bnozss}) with
$\Rm\gg1$, where $\RU$ is the turbulent magnetic Reynolds number of the
outflow and $\Beq$ is the magnetic field strength corresponding to
equipartition with the turbulent energy; $\Beq\simeq5\mkG$ in the Solar
neighbourhood of the Milky Way. Thus, the outflow must be strong enough to
support the dynamo action, $\RU\ga0.1$ according to our results
(Figs~\ref{phplcomb} and \ref{Bevolve}). However, any outflow is removing the
large-scale magnetic field as well, and thus adversely affects the dynamo
action. Hence, the large-scale magnetic field decays when $\RU\ga0.8$. The
strength of the outflow optimal for the dynamo action is $\RU\approx0.3$, and
this is consistent with the plausible range of $\mean{U}_z$ in spiral galaxies
estimated by \citet{SSSB06} as $\mean{U}_z=1$--$2\kms$.

For $\RU=0.3$, the dynamo achieves a steady state in about $10^{10}\yr$ for
parameter values typical of the Solar neighbourhood of the Milky Way, with the
amplification factor of about $10^4$ in terms of magnetic energy. This imposes
significant restrictions on the strength of the seed magnetic field required
for the dynamo \citep{BBMSS96}. However, this estimate, often taken as
representative of spiral galaxies as a whole, only applies to the Solar
vicinity of the Milky Way and, in addition, relies on various poorly known
factors as well as on the approximations of this paper. Closer to the Galactic
centre, the angular velocity of rotation $\Omega\propto r^{-1}$ is larger
together with the dynamo number, $D\propto\alpha\Omega\propto\Omega^2$. For
example, at a galactocentric distance equal to half the Solar orbit radius,
the dynamo number is four times larger than near the Sun, and the dynamo
growth time $\tau\propto (\sqrt{D}-\sqrt{D\crit})^{-1}$ is more than five
times shorter than near the Sun. This applies to other galaxies as well: for
example, the growth rate of the large-scale magnetic field in the nearby
galaxy M51 is estimated to be {\em ten\/} times larger than that in the Solar
neighbourhood \citep[\S{VII.9} in][]{RSS88}.

These features of the mean-field dynamo facilitated by the advective flux of
magnetic helicity appear to be quite satisfactory. However, the steady-state
strength of the large-scale magnetic field, about $0.1\Beq\simeq0.5\mkG$ for
$D=2D\crit$ and $\RU=0.3$ -- see Eq.~(\ref{Bnozss}) and Fig.~\ref{Bevolve}, is
near the lower end of the range observed in spiral galaxies, 1--$5\mkG$
\citep{B00}. The main factor which makes the magnetic field strength low is
$\meanB\propto C^{-1/2}\propto l_0/h\simeq0.2$ in Eq.~(\ref{Bnozss}). The
origin of this factor is the fact that the magnetic helicity evolves over the
time scale $l_0^2/\etat\simeq l_0/v_0$ -- see Eq.~(\ref{fullset2flux}) --
which is shorter than the evolution time scale of the mean magnetic field,
$h^2/\etat$. Thus, even in the presence of the advective helicity flux
magnetic helicity can partially cancel the kinetic one before the large-scale
magnetic field has grown enough. (Without any helicity flux, this cancellation
is more complete and the mean magnetic field is catastrophically quenched.)
The advection term in Eq.~(\ref{fullset2flux}) opposes the growth of
$|\alpha\M|$, and so the steady-state strength of the mean field increases
with $\RU$ as $(\RU/C)^{1/2}$.  We note in this connection that it is
important that Eq.~(\ref{amchi}) involves a scale ($l_0$) independent of the
magnetic Reynolds number; otherwise, the strength of the mean magnetic field
would be catastrophically small for large $\Rm$.

Although the agreement with observations within an order of magnitude may be
sufficient for a crude model explored, we discuss in the Sect.~\ref{vcflux} an
additional effect that can make the magnetic field stronger and accelerate its
growth. We have considered an additional flux of magnetic helicity, suggested
by \citet{VC01}, which arises because of the symmetry breaking and anisotropy
introduced in the turbulent flow by differential rotation. This attractive
mechanism is essentially nonlinear and can only be efficient if the
large-scale magnetic field is strong enough -- in fact, almost exactly as
produced owing to the advective helicity flux (Sect.~\ref{VCdynamo}).
Therefore, we believe that the primary role of the Vishniac-Cho helicity flux
in galactic dynamos is to complement the action of the advective flux of
magnetic helicity. We have not explicitly included another type of magnetic
helicity flux suggested by \citet{KMRS00} and \citet{KMRS02}, which also
relies on anisotropy of turbulence. However, the main elements of the
functional form of that flux are, at least in the approximation used here,
similar to that of the Vishniac-Cho flux, and we believe that the results
would remain qualitatively unchanged.

The additional helicity flux due to anisotropy of turbulence can be essential
for supporting large-scale magnetic fields in starburst galaxies or in
galaxies with a relatively strong wind, where the strong outflow could
otherwise suppress the dynamo action.

Even for $\RU=0$, the mean magnetic field achieves strength of order $0.1\Beq$
before it is catastrophically quenched. As follows from Fig.~\ref{Bevolve},
the field strength is maximum at a time of order $10^{10}\yr$, which is
comparable to the galactic lifetime, and the subsequent decay by a factor of
ten takes another $10^{10}\yr$. This implies that some galaxies can have a
significant mean magnetic field even if their fountain flow is too weak to
alleviate the catastrophic quenching of the mean-field dynamo.

A feature of both the earlier models of dynamical quenching and the present work
is that the turbulent diffusion is assumed to be unaffected
by magnetic field. Due to this reason, in the absence of a helicity flux, as
$\alpha$ goes below a critical value (due to increase in $|\alpha\M|$),
turbulent diffusion leads to a decay of the mean field faster than
at the resistive time-scale, resulting in the rapid decay of the mean field
for $\RU=0$ mentioned in the previous paragraph.
However this is not accompanied in the models by an
equally rapid decay of $\alpha\M$. This is consistent with the fact
that the steady-state solution of Eq.~(\ref{fullset2flux}),
$\alpha\M= \Rm \meanEMF\cdot\meanBB/\Beq^2$ for $\RU=0$, can be of order unity
if $\meanEMF\cdot\meanBB\sim \Rm^{-1}$.
One possibility is that current models are limited in how accurately
they incorporate strict total helicity conservation.
It is also possible that one may have a "turbulent" diffusive 
contribution to the small-scale helicity flux, when one goes to
the next order in large-scale derivatives in Eq.~\ref{finhel}
and the above problem might have resulted from the neglect of this flux.

Alternatively, a mean-field decay without the correspondingly strong
decay in $\alpha\M$, can still be consistent
with strict total helicity conservation if there is a preferential
loss of the mean-field helicity through the boundary
(without the corresponding loss of the small-scale helicity).
Indeed such preferential loss of the large-scale field and its helicity
(apparently due to the turbulent magnetic diffusion of the mean field)
are seen in the direct simulations of
\citet{BD01} where open boundary conditions were used.
In these simulations helical turbulence was driven in a slab between 
open boundaries, with the field was taken to be purely vertical on the boundaries. 
The resulting $\alpha^2$-dynamo led to a mean-field, but with 
a steady-state magnetic energy decreasing with $\Rm$.
This presumably resulted from the fact that turbulent diffusion of
the mean-field (and its helicity) through the boundary was more
efficient than similar turbulent diffusive losses of the small-scale helicity.
Such a behaviour is similar to the catastrophic quenching of the mean field
that obtains in one-dimensional mean field dynamo models (including $\alpha^2$ dynamo 
models) with open boundaries, such as this work or those of \citet{BS05c} and \citet{SSSB06},
in the limit of zero small-scale helicity flux.
Remarkably, in a domain with closed or periodic boundaries, where 
both large- and small- scale helicity losses are zero, a different
type of $\Rm$-dependent evolution obtains. Direct simulations of \citet{B2001} show
that in this case the magnitude of the large-scale magnetic field
can eventually reach larger, super-equipartition  values, 
but the time scale to reach this value 
increases with $\Rm$. 
Clearly, the case of open boundaries is more relevant for galaxies,
and both the direct simulations and mean field dynamo models 
indicate that efficient removal of the small-scale magnetic
helicity from the dynamo region is necessary for a healthy mean-field dynamo action.

On the technical side, we have confirmed the applicability of
the no-$z$ approximation to the fairly complicated nonlinear
dynamo system discussed here. In particular, it is reassuring that,
the time evolution of the magnetic field in the no-$z$ model follows
closely that obtained in numerical solutions of the corresponding
partial differential equations,
including the oscillatory behaviour  \citep{SSSB06}. A nontrivial issue
in this term approach is how to approximate $\meanJJ\cdot\meanBB$, since
this vanishes in the no-$z$ approximation.
Using a different approximation (Appendix~\ref{appen}), we show that this term
is proportional to $-|D|^{1/2} \meanB_r\meanB_\phi$.
\citet{KMRS02} assume a similar form but without the factor $-|D|^{1/2}$
which noticeably affects the quantitative results (especially because of the
different sign of the term). \citet{SSSB06} approximate this term as being proportional
to $\alpha\M$ in their analytical estimate of the steady-state field, their
Eq.~(10), which affects the result for $\RU=0$. With this form, the time
evolution of the field (in the no-$z$ approximation) is also not
correctly reproduced for $\RU=0$.

The mechanisms of dynamo action discussed here can have extensive implications
for galactic magnetic fields. For example, the modulation of the dynamo action
by the outflow may contribute to the formation of magnetic arms, i.e.,
spiral-shaped regions of strong large-scale magnetic field located between the
gaseous spiral arms, as in the galaxy NGC~6946 \citep[e.g., Sect.~5
in][]{B00}. Magnetic arms occur not in all galaxies, and in some cases (e.g.,
M51) the large-scale magnetic field is stronger in the gaseous arms at some
radii and between them elsewhere. It is reasonable to expect that the
intensity of the galactic fountain, quantified by $\RU$, is higher in the arms
where star formation is more intense. If then the average value of $\RU$ is
less then 0.3, the enhancement of $\RU$ in the gaseous arms will lead to a
stronger magnetic field there. Otherwise, for stronger overall outflow,
enhancement of $\RU$ will suppress the magnetic field in the arms, which can
produce magnetic arms interlaced with the gaseous spiral arms.

\section*{Acknowledgements}
We are grateful to E.~Blackman, A.~Brandenburg and J.~Whitaker for useful comments. AS and KS were supported by the Leverhulme Trust via grant F/00~125/N. 
SS would like to thank Council of Scientific and Industrial Research, India for
financial support. AS gratefully acknowledges partial financial support of the
Royal Astronomical Society.

\appendix

\section{Perturbation solution for the ${\alpha}{\omega}$-dynamo}\label{appen}
The kinematic ${\alpha}{\omega}$-dynamo in a thin disc is governed by the
following equations written in dimensionless form \citep[e.g.,][]{RSS88}:
\EQ
\label{Br}
\gam\meanB_r=-\Ral\deriv{}{z}(\alpha\meanB_\phi)+\deriv{^2 \meanB_r}{z^2}\,,
\EN
\EQ
\label{Bphi}
\gam\meanB_\phi=\Rw\meanB_r+\deriv{^2 \meanB_\phi}{z^2}\,,
\EN
with the vacuum boundary conditions at the disc surface,
\begin{equation}\label{BCq}
\meanB_r|_{z=1}=\meanB_{\phi}|_{z=1}=0\,.
\end{equation}
where $\gamma$ is the growth rate of the magnetic field,
$\partial\mean{\BB}/\partial t=\gamma\mean{\BB}$, and $\meanB_z$ can be
recovered from the solenoidality condition. Here we derive an approximate
solution of this eigenvalue problem for $|D|\equiv|\Ral\Rw|\ll1$. As it often
happens with such asymptotic methods, the solution can be applied even for a
relatively large $|D|$. For the sake of definiteness, we assume $D<0$,
$\alpha(z)=\sin\pi z$ and consider quadrupolar modes that are dominant in a
thin disc.

Using new variables $\meanB_r'=\Ral^{-1}\meanB_r$ and
$\meanB_\phi'=|D|^{-1/2}\meanB_\phi$, we rewrite Eqs~(\ref{Br}) and
(\ref{Bphi}) in terms of the dynamo number and then represented them in a
symmetric form
\EQ \label{perteqn}
\gam{\meanBB}=(\w+\e\vp){\meanBB}\,,
\EN
where $\e=|D|^{1/2}$ is a small parameter and for $D<0$,

\[
\w=
\left(
\begin{array}{cc}
\partial^2/\partial z^2 &0              \\
0               &\partial^2/\partial z^2\\
\end{array}
\right),
\]
\[
\vp \meanBB=
\left(
\begin{array}{cc}
0               &-\partial(\alpha \meanB_\phi)/\partial z\\
-\meanB_r       &0\\
\end{array}
\right),
\]
and we have dropped dash at the newly introduced variables.

The eigensolutions of the unperturbed (free-decay) system $\lambda_n
\bb_n=\w\bb_n$  (with the above boundary conditions) are doubly degenerate and
given by
\EQ \label{eigvalue}
\lambda_n=-\pi^2\left(n+\sfrac12\right)^2,
\qquad
n=0,1,2,\ldots,
\EN
\[
\label{eigfunc}
\bb_n= \left(
\begin{array}{c}
          \sqrt{2}\cos[{\pi}(n+{\frac{1}{2}})z] \\ 0
\end{array} \right),
\]
\[
\bb'_n= \left( \begin{array}{c}
                      0 \\ \sqrt{2}\cos[{\pi}(n+{\frac{1}{2}})z]
\end{array} \right),
\]
where we have normalized them to $\int_0^1\bb_n^2\,dz=\int_0^1\bb_n'^2\,dz=1$;
the eigenfunctions should not be confused with the small-scale magnetic field
denoted $\bb$ in the main text.

The expansions
\EQ
\gam=\gam_0+\e\gam_1+\e^2\gam_2+\ldots\,,
\EN
\EQ
\mean{\BB}=\Co\bb_0+\Cop\bb_0'+{\e}\Cl\bb_1+{\e}\Clp\bb_1'+\ldots
\EN
are substituted into Eqs~(\ref{Br}) and (\ref{Bphi}), terms of like order in
$\e$ collected, the dot product of the resulting equations taken first with
$\bb_n$ and then with $\bb_n'$, and results are integrated over $0\leq
z\leq1$. To the lowest order, this yields $\gamma_0=\lambda_0$. A homogeneous
system of algebraic equations for $\Co$ and $\Cop$ follows from terms of order
$\e$, whose solvability condition yields
$\gamma_1=(V_{0'0}V_{00'})^{1/2}=\sfrac12\sqrt{\pi}$, and
$\Cop=\Co\gamma_1/V_{00'}=-2\Co/\sqrt{\pi},$ where
$V_{nm}=\int_0^1\bb_n\cdot\vp\bb_m\,dz$ are the perturbation matrix elements,
whose direct calculation yields $V_{00}=V_{0'0'}=V_{10}=V_{1'0}=V_{1'0'}=0$,
$V_{00'}=-\pi/4$, $V_{0'0}=-1$ and $V_{10'}=-3\pi/4$. To the same order in
$\e$, we similarly obtain inhomogeneous algebraic equations for $\Cl$ and
$\Clp$, which yield $\Clp=0$ and $\Cl=\Cop
V_{10'}/(\lambda_0-\lambda_1)=3\Co/(4\pi^{3/2})$.

\begin{figure} \begin{center}
\includegraphics[width=1.0\columnwidth,angle=0,
bbllx=28bp,bblly=180bp,bburx=563bp,bbury=718bp]{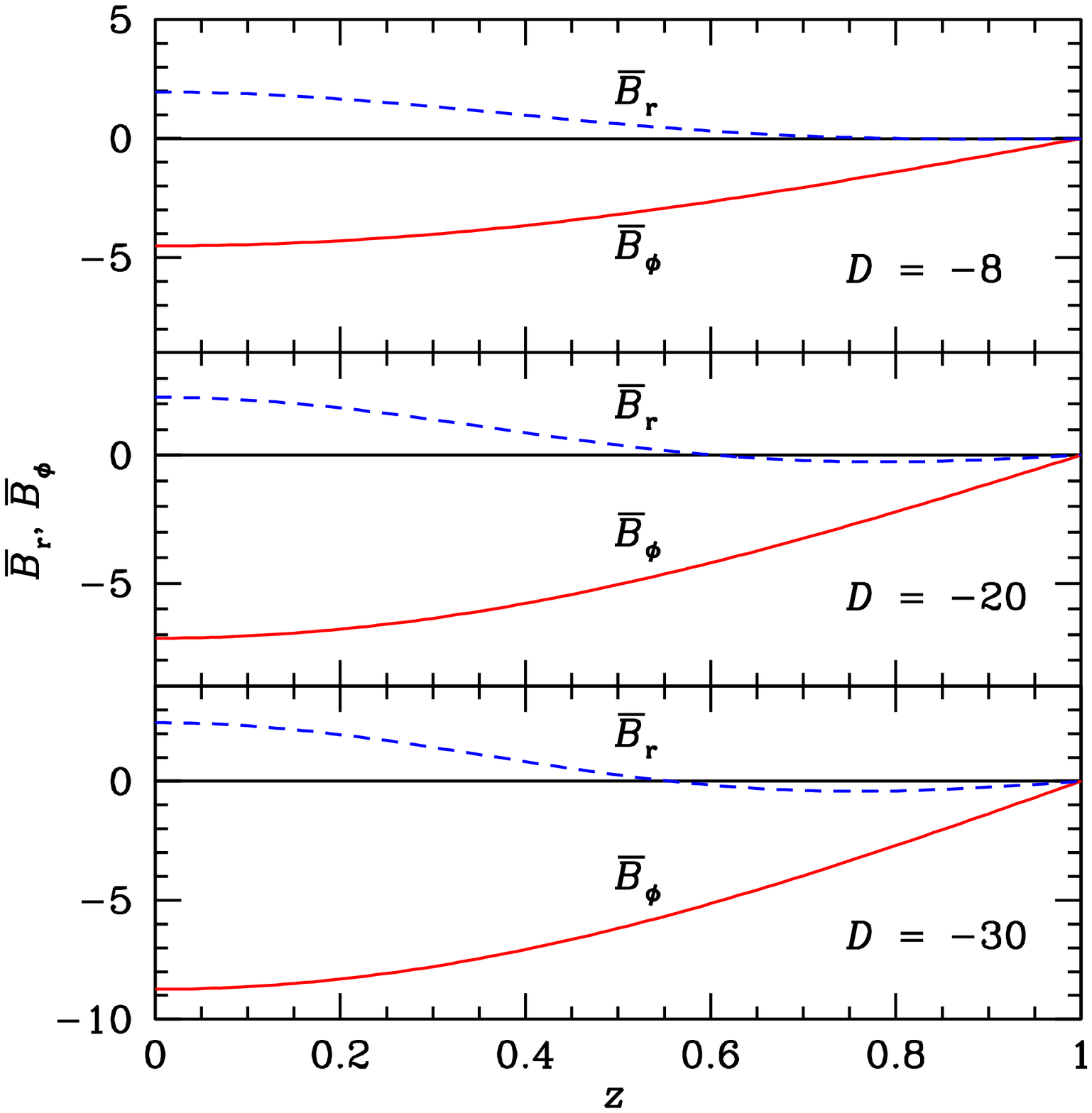}
\end{center}
\caption[]{The approximate eigenfunctions $\meanB_r$ (dashed) and
$\meanB_\phi$ (solid) from
Eqs~(\ref{Br1}) and (\ref{Bphi1}) for $D=-8, -20, -30$ (with
$D\crit\approx-8$), $\Co=1$ and $\Ral=1$.
}
\label{eigfd}
\end{figure}

In terms of the original variables $\meanB_{r,\phi}$, the resulting approximate
solution is given by
\EQ
\label{Br1}
\meanB_r\approx\Ral\Co
\left[
\cos\left({{\pi z}\over 2}\right)
+\frac{3}{4\pi^{3/2}}\sqrt{|D|}\cos\left({{3\pi z}\over 2}\right)\right],
\EN
\EQ
\label{Bphi1}
\meanB_\phi\approx-2\Co\sqrt{\frac{|D|}{\pi}}\cos\left({{\pi z}\over
2}\right),
\EN
\begin{equation}
\gamma\approx -\sfrac14\pi^2+\sfrac12\sqrt{\pi|D|}\,,
\end{equation}
where $\Co$ is an arbitrary constant. Thus, $\gamma>0$ for $D<D\crit$ with
$D\crit\approx-\pi^3/4\approx-8$. This estimate of $D\crit$ is impressively
close to that obtained numerically \citep{RSS88}. We plot the components of
the magnetic field for various values of the dynamo number in
Fig.~\ref{eigfd}; the result reproduces very closely the more accurate
numerical solutions presented, e.g., in Fig.~VII.1 of \citet{RSS88}, including
such a subtle detail as a zero of $\meanB_r$ at $0<z<1$ for $D<D\crit$ which
shifts to smaller $z$ as $|D-D\crit|$ increases.

The current helicity density of this magnetic field follows as
\EQ
\label{JBF}
\mean{\JJ}\cdot\mean{\BB}\approx
-{3\pi^{1/2}\over 8}|D|^{1/2}\meanB_r\meanB_\phi
\EN
to the lowest order in $D$. This estimate is used in Sect.~\ref{noz}.


\label{lastpage}
\end{document}